\documentclass[3p]{elsarticle}
\usepackage[T1,T2A]{fontenc}
\usepackage[utf8]{inputenc}
\usepackage[english]{babel}
\usepackage{lineno,hyperref}
\usepackage{textgreek}
\usepackage{graphicx}
\usepackage{subfig}
\usepackage[dvipsnames]{xcolor}
\usepackage[]{amsmath, amssymb} % Formula subscripts using \ce{}
\usepackage{chemformula}
\usepackage{marginnote}
\modulolinenumbers[5]
\usepackage{soul}
\usepackage{marginnote}
\usepackage{bookmark}
\DeclareUnicodeCharacter{2212}{-}
\DeclareUnicodeCharacter{03C3}{-}
\DeclareUnicodeCharacter{03BC}{-}
\DeclareUnicodeCharacter{03B1}{-}
\DeclareUnicodeCharacter{2500}{-}
\newcommand{\beginsupplement}{%
\setcounter{table}{0}
\renewcommand{\thetable}{S\arabic{table}}%
\setcounter{figure}{0}
\renewcommand{\thefigure}{S\arabic{figure}}%
}

\journal{}

\begin{document}

\begin{frontmatter}
 
\title{Influence of oxygen-defects on intraband terahertz conductivity of carbon nanotubes \iffalse{Influence of plasma-treatment on intraband conductivity in carbon nanotubes}\fi}

%% Group authors per affiliation:

%% or include affiliations in footnotes:
\author[1]{Maksim I. Paukov}
\author[2,3]{Shuang Sun}
\author[4]{Dmitry V. Krasnikov}
\author[1]{Arina V. Radivon}
\author[1,5]{Emil O. Chiglintsev}
\author[1,5]{Stanislav Сolar}
\author[5,6]{Kirill A. Brekhov}
\author[7]{Gennady A. Komandin}
\author[1,8]{Andrey A. Vyshnevyy}
\author[7]{Kirill I.~Zaytsev}
\author[7]{Sergei V.~Garnov}
\author[11]{Nadzeya I.~Valynets} 
\author[4]{Albert G. Nasibulin}
\author[9,8]{Aleksey V. Arsenin}
\author[8]{Valentyn Volkov}
\author[1,5]{Alexander I. Chernov}
\author[2,3]{Yan Zhang}
\author[1,7,10]{Maria G. Burdanova*}

\address[1]{Moscow Center for Advanced Studies, 20 Kulakova St., Moscow, 123592, Russia}

\address[2]{Beijing Key Laboratory for Metamaterials and Devices, Key Laboratory of Terahertz Optoelectronics, Ministry of Education of China,  Beijing, 100048, China}

\address[3]{Beijing Advanced Innovation Center for Imaging Technology, Department of Physics, Capital Normal University, Beijing, China}

\address[4]{Skolkovo Institute of Science and Technology, 3 Nobel St., Moscow, 121205, Russia}
\address[7]{Prokhorov General Physics Institute of the Russian Academy of Sciences, 38 Vavilov St., Moscow, 119991, Russia}
\address[5]{Russian Quantum Center, 30 Bolshoi Boulevard, Moscow, 121205, Russia}
\address[6]{MIREA─Russian Technological University, 78 Vernadsky Avenue, Moscow, 119454, Russia}
\address[9]{Laboratory of Advanced Functional Materials, Yerevan State University, Yerevan, 0025, Armenia}
\address[8]{Emerging Technologies Research Center, XPANCEO, Dubai, United Arab Emirates}
\address[11]{Research Institute for Nuclear Problems of Belarusian State University, Minsk, Belarus}
\address[10]{Osipyan Institute of Solid State Physics of the Russian Academy of Sciences, 2 Academician Osipyan St., Chernogolovka, Moscow region, 142432, Russia}

*Corresponding author: burdanova.mg@gmail.com

\begin{abstract}

The exceptional charge transport properties of single-walled carbon nanotubes (SWCNTs) enable numerous ultrafast optoelectronic applications. Modifying SWCNTs by introducing defects significantly impacts the performance of nanotube-based devices, making defect characterization crucial. This research tracked these effects in oxygen plasma-treated SWCNT thin films. Sub-picosecond electric fields of varying strengths and additional photoexcitation were used to assess how defects influence charge carrier transport. Changes in effective conductivity within the terahertz (THz) range were found to be strongly dependent on impurity levels. The plasmon resonance shift to higher THz frequencies aligns with the defect-induced reduction in conductivity and slowed carrier migration within the network. An increase in THz field strength resulted in diminished conductivity due to intraband absorption bleaching. To address the emergence of hot charge carriers, a modified Drude model, which considers non-equilibrium charge carrier distribution via field-dependent scattering rates, was applied. The dominant charge-impurity scattering rate in plasma-treated samples corresponded with an increase in defects. Additionally, the impact of defects on charge carrier dynamics on a picosecond timescale was examined. The modeled plasma-treated SWCNTs wire-grid polarizer for the THz range reveals the potential for multi-level engineering of THz devices to customize properties through controlled defect populations.

\end{abstract}

\begin{keyword}
single-walled carbon nanotubes, defects, hot-carriers, conductivity, non-linear spectroscopy, polarizers
\end{keyword}

\end{frontmatter}

\section{Introduction}

Carbon nanotubes (CNTs) belong to the rich family of carbon nanomaterials, which are in high demand for the new generation of terahertz (THz) optoelectronics \cite{Zhou2023, Burdanova2021a}. This intense interest in CNTs is driven by their properties, such as high and tunable THz conductivity \cite{Zhang2013, Wang2024,He2019,Chernov2009}, anisotropic properties \cite{Jeon2002}, \cite{Voronin2024}, \cite{Ermolaev2023}, giant polarization rotations \cite{Baydin2021}, and negative photoconductivity \cite{Burdanova2019}, to name a few. These features, in turn, arise from the possibility to tailor CNT characteristics via the changes at different level of organization: from atomic structure through bundle organization and morphology to patterning \cite{Burdanova2021}. This allowed to propose and develop a variety of the optically active CNT-based devices: polarizers \cite{Xu2018}, modulators \cite{Paukov2023}, lenses \cite{Katyba2023}, vortex generators \cite{Radivon2024}, diffraction gratings \cite{Novikov2025} and etc \cite{Toikka2024}.

Being a promising material for THz applications, CNTs require a credible characterization of defects created during the synthesis or postprocessing step, because they can significantly influence the performance of a device \cite{Banerjee2020}. When optimized, this impact might be positive, as defect sites facilitate chemical functionalization \cite{DelCanto2010}, which can expand SWCNTs' potential for use in electronics, energy storage, and catalysis \cite{Hirsch2005} or modify the bandgap of carbon nanotubes by creating localized states within this bandgap \cite{Gifford2020, He_2017}, leading to enhanced electrical properties for various applications, such as sensors \cite{Zanolli2009, Chernov_2017} and other devices. Lastly, the introduction of defects as well as related functional groups allows to tune the SWCNT spectrum in the infrared region allowing advances in, for example, bolometers \cite{Kurtukova2023}. On contrary, defects in the crystal lattice of carbon nanotubes can disrupt the flow of electrons \cite{Reyes2009}, reducing their ability to conduct electricity efficiently, suppress photoluminescence and degrade their mechanical properties making them easier to break or deform under stress.

The recently developed techniques of linear and non-linear THz spectroscopy pave the way for a comprehensive understanding of the fundamental properties of CNTs \cite{Zheng2022, Lauret2003}. The conductivity of SWCNTs at THz frequencies at low fields is primarily explained by the Drude--Lorentz model. This model combines a Drude--type free--carrier response, which describes the movement of charges over long intertube distances throughout the nanotube network. Additionally, it includes a plasmon--type resonance, which accounts for the collective contribution of charges confined to individual nanotubes \cite{Zhang2013, Karlsen2017}. The plasmon resonance frequency depends on the localization distance determined by the nearest ends, impenetrable defects, and intertube contacts. Meanwhile, the thermalization time of accelerated carriers is short in CNTs \cite{Kar2019}, similar to graphene \cite{Johannsen2013}, and the nonlinearities created by a high-strength field can be quantified as an elevated temperature of the electronic subsystem \cite{Kar2019}. Low- and high---field THz, as well as optical pump-THz probe spectroscopy, are capable of indicating the influence of defects on AC conductivity and photoconductivity, and this idea has not yet been thoroughly utilized. Meanwhile, several related results have been reported for graphene \cite{Docherty2012}. 

In the following, we present a qualitative systematic study on the influence of defects induced by oxygen plasma on the electronic transport of SWCNTs, employing low-- and high--field strength time--domain THz as well as optical pump -- THz probe spectroscopy. We tracked the conductivity changes over a broad range of frequencies and applied electric field strengths. The conductivity of SWCNTs at high applied fields can be described using the Drude--like transport equation with an energy--dependent scattering rate. The combination of analytical methods, such as absorption and Raman spectroscopy, along with high--field THz spectroscopy, allows us to qualitatively track the changes in short--range scattering with the increase of oxygen defects across the set of samples. In addition, we show the change in the carrier relaxation dynamic under femtosecond laser irradiation at 400\,nm. Finally, we present a finite element model of a THz wire grid polarizer based on the obtained dielectric characteristics for defective samples and various field strengths. The proposed modeling and recommendations can be utilized to design THz polarizers using SWCNTs with controllably varied defects concentration.

\section{Materials and methods}

Thin SWCNT films were synthesized by the aerosol CVD method based on carbon monoxide disproportionation (the Boudouard reaction) on the surface of a Fe--based aerosol catalyst nanoparticle \cite{Khabushev2019}, \cite{Khabushev2020}. The films contained ca. one--third of metallic and two--thirds of semiconducting SWCNTs due to the random folding \cite{Nasibulin2005}. The average diameter of a tube in the film is estimated to be 2 nm (from the absorbance bands and the Kataura plot). One obtained uniform film was divided into five equal parts for feather press dry transfer to five independent substrates. One of the resulted films was left pristine, while the others were treated with an oxygen plasma for 10, 30, 60, and 90 seconds, respectively. The plasma treatment procedure (Diener Electronic PiCO SL, 0.3 mbar of 99.9999\% O$_2$, ~100 W) was described elsewhere \cite{Kurtukova2023}. Notably, by using the parts of the same film and progressively increasing the plasma treatment duration, we were able to highlight the influence of the defect concentration via the spectroscopic changes.

For structural studies of SWCNT films, we performed Raman spectroscopy (Labram, Jobin-Yvon Horiba) at an excitation wavelength of 532 nm with a spectral resolution of 0.5\,cm$^{-1}$. A UV/vis/NIR (250--800\,nm wavelength) spectrophotometer (Cary) and a Bruker IFS113v IR Fourier spectrometer were used to acquire transmission spectra in the UV/vis/NIR and infrared (30--5000 cm$^{-1}$) ranges, respectively.
The absorbance $A$($\lambda$) was calculated from the transmission $T$ as follows $A=-\log_{10}T$. The spectra of transmission in the infrared range (30–5000 cm$^{-1}$) were recorded by the Bruker IFS113v IR Fourier spectrometer.

Raman spectra were obtained using a confocal scanning Raman microscope Horiba LabRAM HR Evolution (HORIBA Ltd., Kyoto, Japan) with a 532~nm laser wavelength (with a spectral resolution of 0.5~cm$^{-1}$). Measurements were carried out using linearly polarized excitation at wavelength 532~nm, 1800~lines/mm diffraction grating, and $\times$10 objective. The spot size was ~0.4~$\mathrm{\mu}$m. The sample was mounted on motorized stages which allowed for a 2D x--y scan.

Linear (weak-field) transmission THz spectra were collected by the home-made time-domain spectroscopy (TDS) setup \cite{Komandin2019}. THz radiation was generated and detected by photoconductive antennas (Fraunhofer IPM) excited by femtosecond pulses of a fiber laser (Toptica FemtoFErb 780) at a wavelength of 780\,nm and with a pulse repetition rate of 100 MHz. Generated THz pulses had a broad spectral range 0.1--4\,THz reduced to 0.3--3\,THz due to the losses. All measurements were performed in a vacuum chamber.

High--field THz TDS was performed using the femtosecond Yb--doped diode-pumped solid--state amplifier system operating at a wavelength of 1030\,nm with a pulse duration of 30\,fs and a repetition rate of 1\,kHz \cite{Chiglintsev}. The THz radiation with 1.25\,ps pulse length, 2.5\,THz bandwidth, 520\,\textmu W peak power, 900\,\textmu m diameter, and strengths up to 120\,kV/cm was produced by a BNA organic crystal. THz signal was detected by a 1--mm--thick ZnTe crystal. We excluded nonlinear effects in the detection of a signal by checking THz pulse temporal form and spectrum dependencies on the optical pump power.

Complex photoconductivity was measured by a home--made time--domain THz spectroscopy setup in the transmission geometry using the optical pump THz probe (OPTP) technique. The incident laser pulse was generated by a Spectra-Physics femtosecond laser amplifier, seeded with a 1\,kHz, 35\,fs, 800\,nm pulses generated by a Spectra--Physics Mai Tai oscillator. The THz probe pulse was generated using a ⟨110⟩ ZnTe crystal with a thickness of 1\,mm and detected with another ⟨110⟩ ZnTe crystal with a thickness of 1\,mm. The THz beam was focused at the sample to a spot with a diameter of just 1\,mm by the high numerical aperture off-axis parabolic mirrors, while the optical pump beam of 400 nm (the frequency is doubled) had a spot of the order larger --- about 5\,mm for all measurements. The optical pump beam was chopped at 130\,Hz.

Based on the time--domain measurements obtained on all mentioned setups, we can employ the Tinkham thin film equation to calculate the complex conductivity spectra of the SWCNT films, owing to the high absorption characteristics of thin films \cite{LloydHughes2012, Joyce2016}.

\section{Results}

\begin{figure*}[t!]
  \centering\includegraphics[width=0.7\columnwidth]{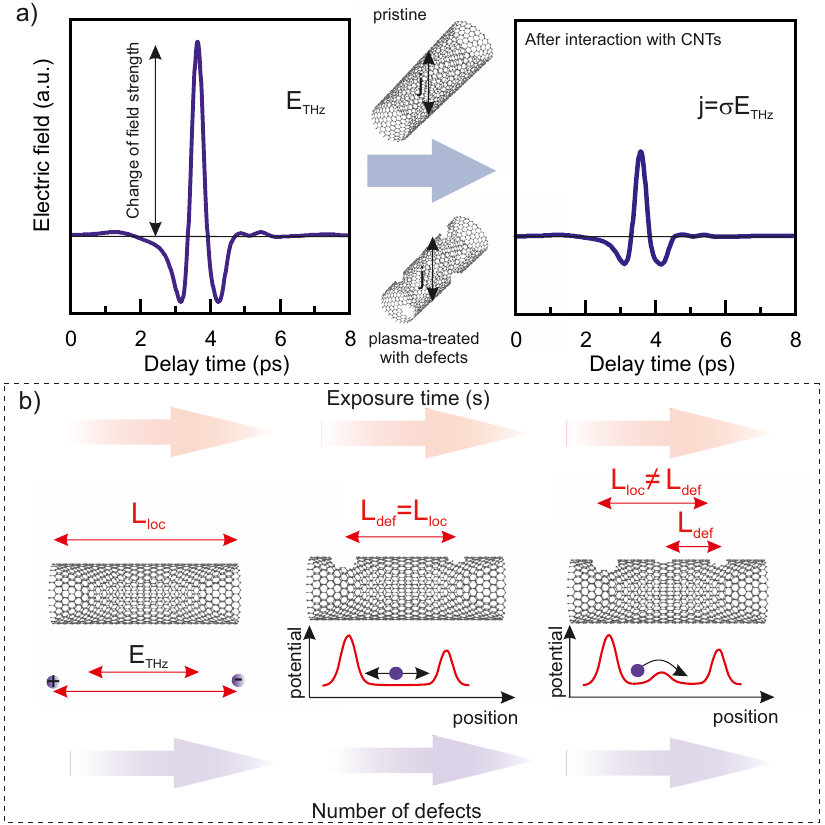}
  \caption{a) The THz pulse applied to SWCNTs induces a current $j$, which emits an additional field proportional to the time--derivative of the initial field. The superposition of the initial and emitted fields results in the reduction of the resultant pulse, which can be backtracked to determine the conductivity of the material. The influence of field strength and defects can be examined. b) Illustration of localization of a charge carrier between two barriers in one dimension. The potential barriers are induced by defects or impurities, and the localization length of the electron wavefunction roughly corresponds to the distance between the barriers. Such localization and the resultant conductivity strongly depend on the height of the potential barriers, as some electrons cannot "hop" over certain barriers to travel from the start to the end of the individual SWCNT. }
  \label{FIG:Fig1}
\end{figure*}

To study the influence of defects on charge transport, we used a combination of linear and non--linear ultrafast THz, as well as optical pump--THz probe spectroscopy. In these techniques, an oscillating electric field with varying strengths applied to a SWCNT network (within a thin film) induces a current proportional to the conductivity (Fig. \ref{FIG:Fig1}a). Defects in one--dimensional nanomaterials have a greater influence on charge transport than in three--dimensional solids. In the 1D case, defects can lead to the confinement of the electron wave function between these defects. This localization significantly decreases the conductivity of the nanotubes, as the conduction electrons have to `hop' over all barriers and shift the plasmon resonance.

\begin{figure*}[t!]  \centering\includegraphics[width=1.0\columnwidth]{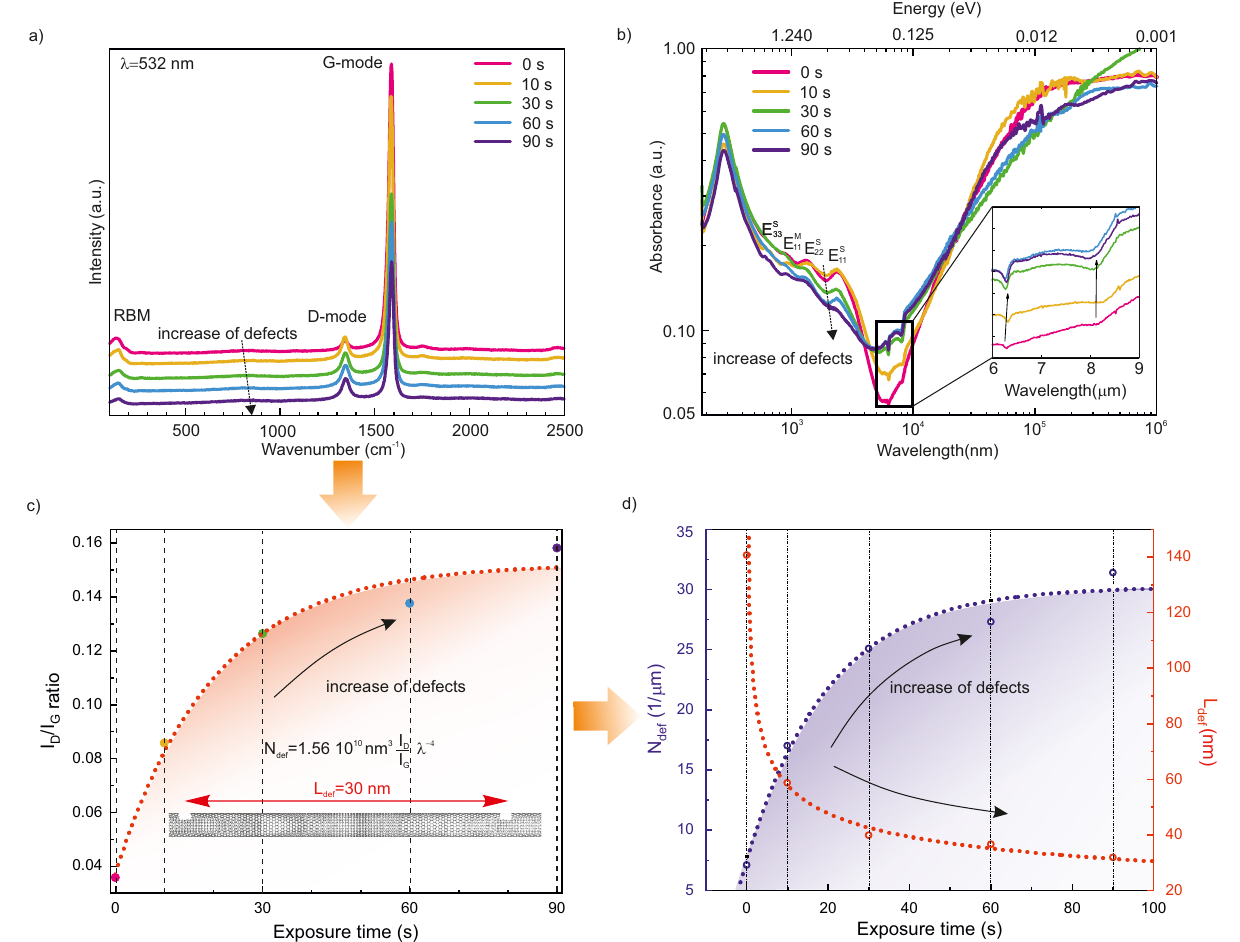}
  \caption{Characterization of the plasma-treated SWCNT thin films. The Raman (a) and absorbance (b) spectra of pristine (pink) and plasma--treated samples of SWCNTs with exposure times of 10, 30, 60 and 90\,s. c) The gradual increase in the Raman peak intensity ratio $I_{D}/I_{G}$ with the change of the plasma exposure time. The inserts show the empirical expression for the calculation of the defect density and the illustration of the distance scale between nearest defects \cite{Dresselhaus1995}. d) The calculated number of defects (left axis) and distance between the nearest defects (right axis) as a function of plasma treatment time. The dashed lines are guides for the eye.}
  \label{FIG:Fig2}
\end{figure*}

Under the applied oscillating electric field, electrons, and holes are driven in opposite directions within individual SWCNTs. As a result, the positive charge is confined to one side of the SWCNTs, while the negative charge accumulates on the opposite side. The electric field generated by this dipole opposes the initial field. Thus, the depolarization field acts as an electrostatic restoring force, causing the carriers to undergo collective harmonic oscillations known as plasmons. These oscillations are present regardless of a particular conductive behaviour of a carbon nanotube. The plasmon frequency theoretically depends on the length or localization distance $L_\mathrm{loc}$ as given by the equation: $\omega=V_{q}(\pi/L_\mathrm{loc})$, where $V_{q}$ is the mode velocity which is proportional to the Fermi velocity and the diameter of the SWCNTs. Moreover, the localization distance does not necessarily need to be equal to the distance between the nearest defects (which will be estimated further from the THz spectroscopy in the text), as some electrons can still 'hop' over certain potential barriers.

The ratio between the intensities of D and G modes of Raman spectra is widely used as a versatile assessment of the SWCNT quality. G mode is usually associated with the tangential stretching mode of the sp$^2$ carbon lattice, while the D and 2D modes, as a rule, are attributed to breathing of carbon hexagon. For the ideal sp$^2$-carbon lattice, two-phonon process (2D) is Raman-active, while the first-order Raman scattering (D mode; ~ 1350 cm$^{-1}$) is prohibited due to the selection rules \cite{Jorio2021}. Nevertheless, any perturbation (i.e. a defect of any kind) in the ideal graphene lattice allows the D mode. This is why $I_D/I_G$ \cite{Canado2006, Dresselhaus2010} or $I_{2D}/I_D$ ratios are commonly used for the quantitative assessment \cite{Kuznetsov2014} of the defective structure. Previous studies of Raman scattering in SWCNTs have determined that the number of defects $n_D$ can be estimated from $I_D/I_G$ according to the relationship, which is proportional to the distance between defects $L_\mathrm{def}$ (see further estimation of this length from Raman measurements in the paper). Therefore, the comparison of $L_{loc}$ and $L_\mathrm{def}$ obtained from the combination of Raman and THz spectroscopy can give a more qualitative understanding of the influence of defects in charge transport(Fig.~\ref{FIG:Fig1}b). 

\subsection{Raman and absorption spectroscopy}

To evaluate the defect density as a function of oxygen plasma exposure time, we performed Raman spectroscopy at an excitation wavelength of 532\,nm. The pronounced peaks of the dominant radial breathing mode (RBM) and G Raman--active modes of the SWCNTs, as well as the defect--related D mode, are clearly observed in the spectra (Fig.~\ref{FIG:Fig2}a). We observe a decrease in the RBM intensities and a slight shift of its position towards higher wavenumbers with the increase in defect density. This behavior is within our expectations and is explained by the change in the number and vibrational nature of the Raman--active modes caused by the defects \cite{Saidi2013}. Additionally, the shift may be caused by oxygen doping that occurs during plasma treatment \cite{Guo2007}.

The defect--related D and G mode also change as the defect density increases. Although the G mode shift is almost undetectable (within the spectral resolution of the measurement setup), the D peak shift reaches as high as 10\,cm$^{-1}$. The main indicator of the defect density is the intensity ratio of D and G modes which visibly grows with the increase in the plasma exposure time (Fig.~\ref{FIG:Fig2}a,c). Similar behavior was also observed in graphene \cite{Dresselhaus2010}. Notably, when the defect density in graphene reaches a certain value, its crystalline structure becomes disordered, and a further increase in the defect density leads to a decrease in the D-to-G intensity ratio instead. In our case, the monotonic behavior of $I_D/I_G$ shows that we do not reach the saturation with a following significant drop, thus for the defect density estimation we employed empirical expression: $n_\mathrm{d}=1.59\times10^{10}\mathrm{nm}^3\frac{I_D}{I_G}\lambda^{-4}$, where $\lambda=532$\,nm is the excitation wavelength in the Raman experiment\cite{Wang2017}. From the defect density we find the average distance between nearest defects as $L_\mathrm{def} = 1/n_\mathrm{d}$ (see Fig.~\ref{FIG:Fig2}d). At the maximum oxygen plasma exposure time of 90\,s, it is still as low as 30\,nm which is about 4 times smaller than that for the pristine sample. Here we again draw the readers attention that this is not a direct measurement of the distance between defects, it is rather the effective parameter.

Broad-band optical absorbance in the visible range demonstrates a series of peaks, which are attributed to the exciton transitions between the corresponding Van Hove singularities of metallic and semiconducting SWCNTs (see the denoted peaks $\mathrm{E}_{ii}^{S,M}$ in Fig.~\ref{FIG:Fig2}b). The corresponding values of the resonant wavelengths are 2392 ($\mathrm{E}_{11}^S$), 1338 ($\mathrm{E}_{22}^S$), 883 ($\mathrm{E}_{11}^M$), and 618 ($\mathrm{E}_{33}^S$)\,nm. With the increase in defect concentration, $\mathrm{E}_{11}^S$ and $\mathrm{E}_{22}^S$ transitions are blue--shifted, whereas $\mathrm{E}_{11}^M$ is red--shifted compared to the pristine sample. 
Defects tend to introduce new mid--gap states, which lead to the redshifts in the lower--energy excitonic feather E$^{S}_{11}$ \cite{Ma2014}. Meanwhile, oxygen doping results in the blueshift of higher--energy E$^{S}_{22}$ and consequent suppression of the excitonic transition due to the Pauli-blocking effect \cite{Kim2008}. The presence of defects also causes excitons to become more confined (i.e., localized), resulting in the broadening of excitonic peaks.

Studying SWCNTs in the mid--IR range allows for the detection of several bands in the spectrum of SWCNTs (see the inset of Fig.~\ref{FIG:Fig1}b). These absorption bands are attributed either to the vibrational modes of SWCNTs, which exhibit a dynamic dipole moment, or to the various stretching modes of carbon atoms and their functional groups \cite{Kurtukova2023}. These features have been observed in both experimental and computational studies \cite{Bantignies2006}. One such band, located at approximately 8.1 $\mu$m, is associated with the overlap between the IR--active vibrational mode of widely distributed SWCNTs, as predicted by modeling \cite{Sbai2006}, and the emergent D--band, which has been confirmed by Raman measurements. The presence of this band becomes more significant with the introduction of defects. However, it is very broad and does not effectively characterize the degree of 'defectiveness' of the structure. The second band, observed at approximately 6.25 $\mu$m, corresponds to the stretching mode of the \ch{C=O} bond (evidenced by its gradual increase). This mode is narrower, allowing one to track not only its height but also its central position, depending on the amount of defects. It is interesting to note that these bands correspond to bleaching in the absorption spectra. This effect has already been observed in the near-infrared range in the presence of O$_2$ \cite{Zheng2020}, but there has been no discussion of middle--infrared features in the literature. Therefore, the origins of this behavior remain elusive.

\begin{figure*}[t!]
  \centering\includegraphics[scale=0.5]{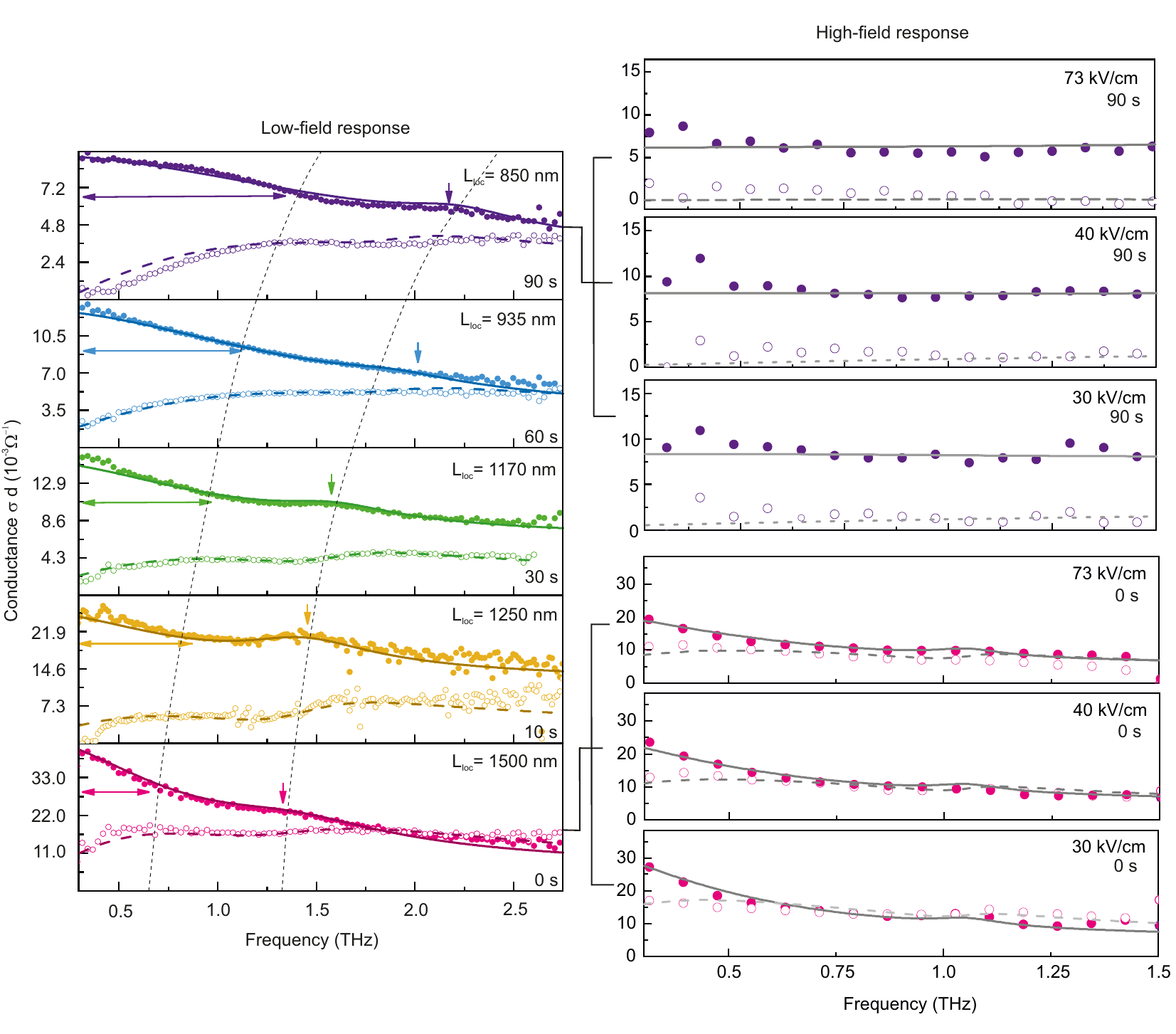}
  \caption{~Frequency-dependent conductivity of the studied samples at low and high field--strength (colors correspond to different plasma exposures similaraly to Fig.~\ref{FIG:Fig1}). Experimental data on the real part of the conductivity is shown with filled dots, the imaginary part -- with empty ones. Solid and dashed curves are fits of the real and imaginary parts of conductivity according to the eq. 1. Colored arrows denote changes in the width of the Drude peak (horizontal arrows) and position of the axial plasmon frequency (vertical arrows). The plasmonic contribution set unchanged with the field strength change while varying between the samples with different plasma treatment times.}
  \label{FIG:Fig3}
\end{figure*}

\begin{figure*}[t!]
  \centering\includegraphics[scale=0.6]{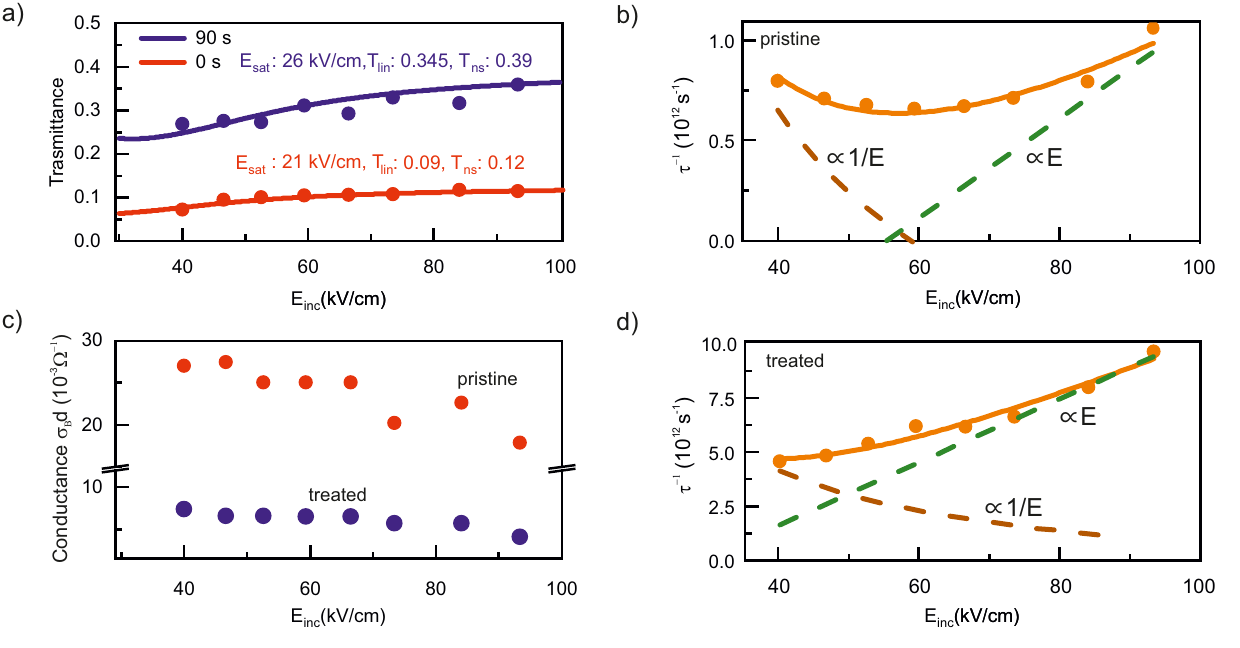}
  \caption{(a)~THz power transmission fitted by the eq. 5. The scattering time for pristine (b) and plasma-treated defective SWCNTs (90 s) (d) versus the electric field. The contribution from short- (brown) and long-range scattering (green) is highlighted by dash lines. (c) The change of the conductance versus electric field strength for pristine (red) and defective nanotubes (navy).}
  \label{FIG:Fig4}
\end{figure*}

\subsection{Terahertz spectroscopy in low-field regime}

The extracted effective complex sheet conductivity at the THz range in the low--field regime is presented in Fig.~\ref{FIG:Fig3}a. The corresponding fit of the Drude--like and Lorentz models, which accounts for the contributions of delocalized free charge carrier response and defect--tunable plasmon resonance of the carriers in finite--length metallic and doped--semiconducting nanotubes with an extra term of the Mott hopping conductivity \cite{Gorshunov2018}, is applied: 
\begin{equation}\label{eq:DLH}
    \sigma(\omega)=\sigma_{\mathrm{D}}\frac{i\gamma_{\mathrm{D}}}{i\gamma_{\mathrm{D}}+\omega}+\sigma_{\mathrm{pl}}\frac{i\omega\gamma_{\mathrm{pl}}}{i\gamma_{\mathrm{p}}\omega+\omega^2-\omega_0^2}+(\sigma_{\mathrm{H}}+A\omega^s),
\end{equation}
\noindent where the first term describes the Drude conductivity of free charge carriers ($\sigma_{\mathrm{D}}$ is the dc Drude conductivity, $\gamma_{\mathrm{D}}$ is the scattering rate), the second - the Lorentz profile of the surface plasmons ($\sigma_{\mathrm{pl}}$ is the plasmon conductivity at the resonance frequency $\omega_0$, $\gamma_{\mathrm{pl}}$ is the scattering rate), and the last one is the Mott hopping conductivity at the junctions with constants $\sigma_{\mathrm{H}}$, $A$, $s$ ($0.5\leq s\leq 1$, describing percolation of the system). The fitting parameters are summarized in Fig.~\ref{FIG:SIFig1}a-c.

Following the discussion of the plasmon frequency change due to the introduction of defects, we estimated the conductivity pathways or localization lengths using the obtained from the fitting parameters of $\omega_0$ (Fig.~\ref{FIG:SIFig1}c). The pristine sample, having an $I_D/I_G$ ratio of 0.04, indicates a low number of defects, and the estimated localization length was 1500\,nm. Due to the low number of defects, this localization primarily results from the potential barriers created at intertube contacts and the ends of individual nanotubes. Plasma--induced defects shorten the conductivity pathways, resulting in a pronounced shift of the resonance position towards higher frequencies and shorter localization lengths, ranging from 1250 to 850\,nm for samples treated with plasma for 10, 30, 60, and 90\,seconds.

Defects also slow down the overall carrier migration within the network, leading to a broadening of the conductivity spectrum and a consequent increase in the electron scattering time(Fig.~\ref{FIG:SIFig1}b). Interestingly, the Raman and THz spectroscopy measurements showed differing changes in the localization distance, which can be explained as follows: some fraction of electrons can still hop over the potential barriers (THz spectroscopy probed) created by sufficiently small defects (probed by Raman spectroscopy). DRough estimate of mean hopping probability is given by ratio of average defect distance ($L_\mathrm{def}$) to effective electron localization length ($L_\mathrm{loc}$). For the pristine sample, this probability is 8.6\% and decreases to up to 3.5\% for the defective SWCNTs samples, which might indicate the non-uniform formation of defects that tend to cluster at the edges of existing defects, expanding them. 

\subsection{Terahertz spectroscopy in high--field regime}

\if
In addition, THz field only induces intra-band transitions and the total number of carries remain constant: 

\begin{equation}
n=\int_{0}^{\infty}g(\varepsilon)f(\varepsilon, T_e)d \varepsilon=const,
\end{equation}

\noindent where $E_F$ is the Fermi energy. 

\fi

Now we turn to the discussion of the conductivity of SWCNTs in a high--field regime (up to 100\,kV/cm). The overview of the experimental results is shown in Fig.~\ref{FIG:Fig3}(low-field spectra panel a and hight-field spectra panel b). We observe a general reduction in the conductivity with an increase in the field strength, similarly to previous reports for graphene \cite{Mics2015}. A high electric field results in modification of electronic temperature, which leads to a change in electronic distribution and Fermi-Dirac electron distribution. To satisfy the principle of conservation of charges, an increase in the electronic temperature must be offset by a decrease in the chemical potential. This relationship provides a rationale for the observed reduction in conductivity with an increased applied field. A similar effect related to charge energy conservation has been observed in the case of doped SWCNTs \cite{Burdanova2019}. Doping induces a shift in the Fermi level, which in turn blocks intraband absorption due to the Pauli--blocking effect, while simultaneously enhancing interband absorption in the THz region. \if This dual effect clarifies why conductivity changes under different doping conditions and motivates the adoption of a thermodynamic approach, rather than a microscopic one, to explain our experimental results.  Additionally, it is important to note that the plasmon contribution remains constant, further supporting our thermodynamic perspective.\fi

To investigate the impact of defect formation under plasma treatment on conductivity in high-strength THz field, we measured complex conductivity for pristine and highly treated samples. To  reproduce the complex conductivity  presented in Fig.~\ref{FIG:Fig3}(right), we varied the Drude component leaving the Lorentz and hopping part unchanged following eq.\,(1). Interestingly, the scattering rate for both pristine and treated samples  exhibits a complex behavior with the increase of the field strength following equation $\frac{1}{\tau} = A\cdot E+\frac{B}{E}$, where A and B are constants. This a result of complex transient behavior of electrons and holes in SWCNT which involves: ultrafast intraband relaxation by e-e collisions \cite{Hirtschulz2008}, strongly asymmetric distribution of carriers in the momentum space, activation of additional channels of momentum relaxation via higher-order subbands. Rigorous theoretical treatment of these processes is a topic of future works.  

Fig\,\ref{FIG:Fig4}a shows the variation of the transmitted signal over the applied electric field strength, ranging from 40 to 93\,kV/cm at about 1 THz. The electric field induced a non-linear response in both pristine (0\,s) and a defective SWCNTs sample (90\,s). The saturable behaviour can be described by a phenomenological model as follows\cite{Hwang2013}:

\begin{equation}
T(E_\mathrm{inc})=T_\mathrm{ns}\frac{\mathrm{ln}\big[1+\frac{T_\mathrm{in}}{T_\mathrm{ns}}(\exp({E^{2}_\mathrm{inc}/{E^{2}_\mathrm{sat}}})-1)\big]}{E^{2}_\mathrm{inc}/{E^{2}_\mathrm{sat}}}, 
\end{equation}

\noindent where $E_\mathrm{inc}$, $E_\mathrm{sat}$, $T_\mathrm{lin}$, and $T_\mathrm{ns}$ are the incident peak electric field, saturable electric field, the linear power transmission, and no-saturable transmission, respectively. Saturation in power transmission is observed at electric fields of 21\,kV/cm for pristine SWCNT and 26\,kV/cm for defected SWCNTs. $T_\mathrm{lin}$ and $T_\mathrm{ns}$ are higher for plasma-treated SWCNTs due to the lower overall conductivity of defective nanotubes. This saturation phenomenon is explained by a subsequent redistribution of electrons within the conduction band. The pronounced non-linearities exhibited by SWCNTs hold significant potential for applications in THz modulators functioning in a non-linear regime.

\subsection{Optical pump -- THz probe spectroscopy}

\begin{figure*}[t!]
  \centering\includegraphics[scale=0.4]{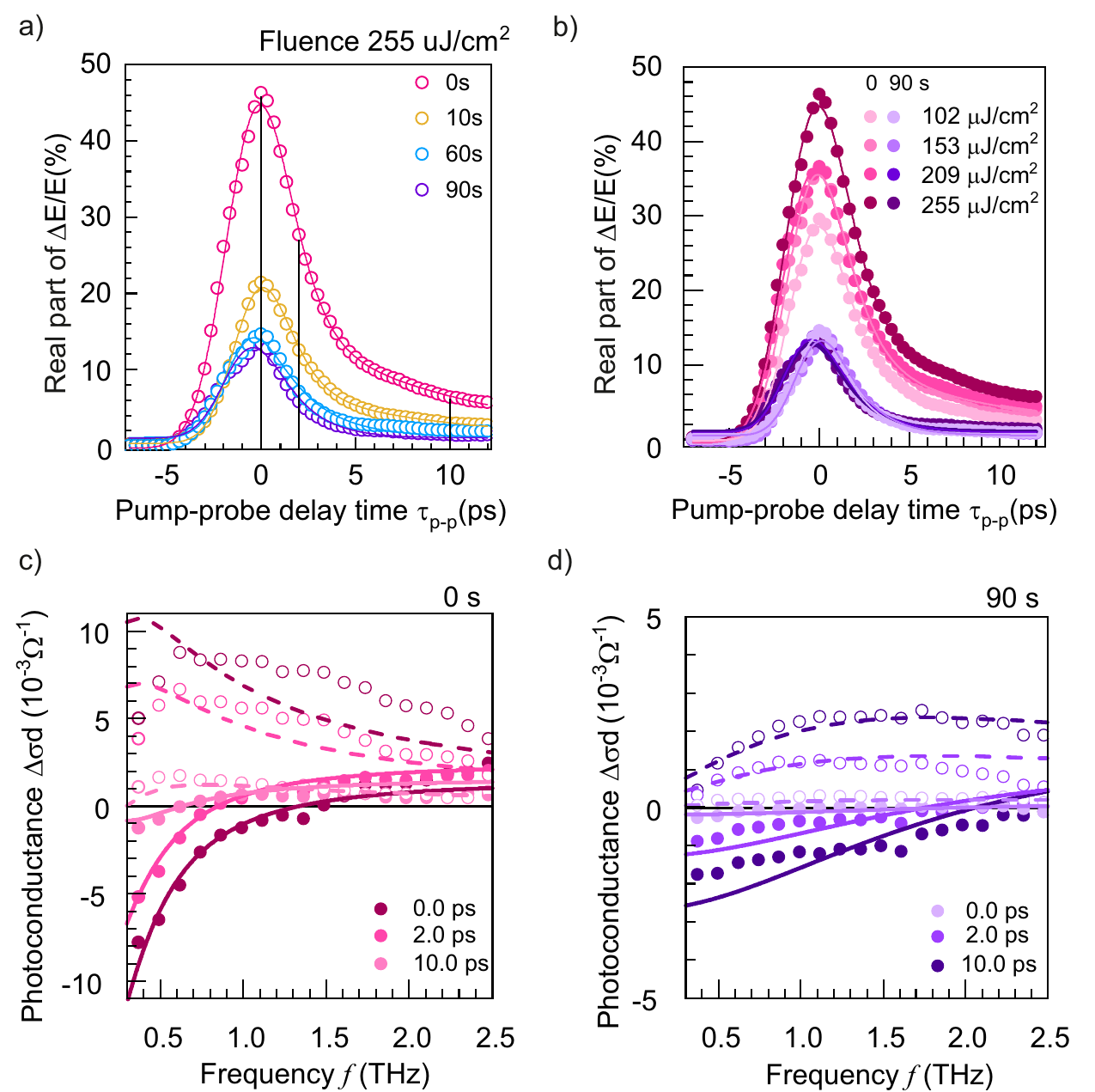}
  \caption{a) The transient of the real part of $\Delta E/E$ for all samples at a fluence 255 $\mu J/cm^2$ and for pristine and treated during 10, 60, and 90\,s by plasma samples. b) The transients for pristine and 90 s plasma-treated sample at different fluences ranging from 102--255 $\mu J/cm^2$. Dots denote experimental values, solid curves are fits to the data, which take a form of the sum of exponential decays convoluted with the Gauss function. Black vertical lines in panel a denote the positions at which the photoconductivity was calculated. The spectra of photoconductance for pristine (c) and plasma-treated during 90 s (d) samples at different pump-probe delay times. Filled and empty dots are the calculated real and imaginary parts extracted from measurements, solid and dashed curves are fits of the real and imaginary parts, correspondingly. }
  \label{FIG:Fig7}
\end{figure*}

In this research, we studied the influence of defects on the dynamical photoconductivity in the THz range. For this purpose, optical pump—THz probe measurements were conducted using 400 nm pump excitation (details are provided in the Methods section). Importantly, 400 nm photoexcitation is off-resonance for the optical transitions of most single--walled carbon nanotubes presented in the sample (see Fig.~\ref{FIG:Fig2}b). The transient change in the transmitted THz field $\Delta E$ due to the optical pump is illustrated in Fig.~\ref{FIG:Fig7}. We observed a twofold reduction in the maximum THz transmission (at a 0 ps pump-probe delay time), which correlated with an increasing number of defects. The most significant change was between the pristine sample and the one treated in oxygen plasma for 10 seconds. In contrast, the change from 10 seconds to 90 seconds of treatment was less pronounced (see Fig.~\ref{FIG:Fig7}a). This phenomenon can be attributed to the increased scattering probability on defects in lightly treated samples as the defect density rises, which explains the observed decrease in photoconductivity. Conversely, in the other samples, the scattering probability becomes saturated, as depicted in Fig.~\ref{FIG:SIFig3}c, which does not influence the overall value of photoconductivity.

\if Taking into account that the pump pulse duration was approximately 50\,fs, we defined that the relaxation dynamics, which were phenomenologically fitted by the form of convoluted Gaussian and exponential decay processes, demonstrated the broadened width of the gaussian function. This width was estimated to be roughly 1.3\,ps regardless of the studied sample, and this phenomenon might be attributed to various phenomena under high optical fluences, such as hot carrier thermalization, phonon bottleneck effect, presence of impurities, and exciton formation\red{[]}.\fi

Fitting of the relaxation dynamics by deconvolution of Gaussian and exponential decay processes revealed that there are two processes governing the decay of the relative change of the THz signal for the pristine sample, and only one for others. The corresponding amplitudes and time constants are shown in Fig. \ref{FIG:SIFig2}a--d. The first fast relaxation process, emerging in all samples in a range of 3\,ps is comparable to previous reports\cite{Perfetti2006, Lauret2003, Kampfrath2005}. Time constant of this process increased almost linearly with the increase of the number of defects. The second process —- previously identified as Auger recombination \cite{Xu2009,Wang2004} -- possessed a higher time constant of 10\,ps and disappeared for samples with defects. \if The high carrier density in SWCNTs leads to strong Coulomb interactions, which provide the possibility of many--body interactions such as the Auger processes. The Auger processes for carrier capture by defects are believed to be possible as well.\fi The analysis of time constants shows that they are independent of fluence (within the error bars), so no nonlinear processes occur in the system in the range of chosen fluences. Since we observed a little fluence dependence in the decay times (Fig.~\ref{FIG:SIFig3}d) of all samples, we ruled out the Auger recombination as a dominant mechanism for all samples. Instead, we associated the THz photoresponse with a cooling of electronic system as described in \cite{Paukov2024}.      
The comparison of the relaxation dynamics for pristine and the mostly plasma-treated samples at different pump fluence are shown in Fig.~\ref{FIG:Fig7}b. The gradual increase of the $\Delta E/E$ ratio for the first sample, and the approximately constant ratio in the other case. 

Before the optical excitation,
electrons in a SWCNT form an electron gas occupying energy bands according to the Fermi--Dirac statistics at the ambient temperature. Upon illumination with an above--bandgap femtosecond laser pulse, free charge carriers are created and lifted up to the respective bands of a SWCNT. These non--equilibrium charge carriers reach thermal equilibrium with the lattice once the extra energy is redistributed and dissipated through different pathways. First, free charge establishes an equilibrium among themselves via elastic carrier--carrier scattering in a
subpicosecond time scale. The photoconductivity contributions of metallic and semiconducting SWCNTs are negative and positive, correspondingly\cite{Paukov2024}. As a result, the spectrum of SWCNT mixture exhibits a peculiar cross--over of the sign of the real part of the photoconductivity for both pristine and plasma--treated samples  (Fig.~\ref{FIG:Fig7}c,d). This can be accounted by the semiclassical transport equation. We note in passing that the semiclassical transport theory transforms into the simple Drude model when scattering rate is energy--independent\cite{Paukov2024}.

It was found out that on the timescale of 10\,ps the relaxation to non--perturbated state is almost achieved, coming to an agreement with the studied dynamics. All samples have the same tendency of the parameters change: after the sharp change from 0 to 2\,ps, the difference appears to be low. Moreover, the samples with defects demonstrate very near changes, from what it might be stated that even a small presence of oxygen defects is high enough to dramatically change the photoconducting parameters.   

\begin{figure*}[t!]
  \centering\includegraphics[scale=0.6]{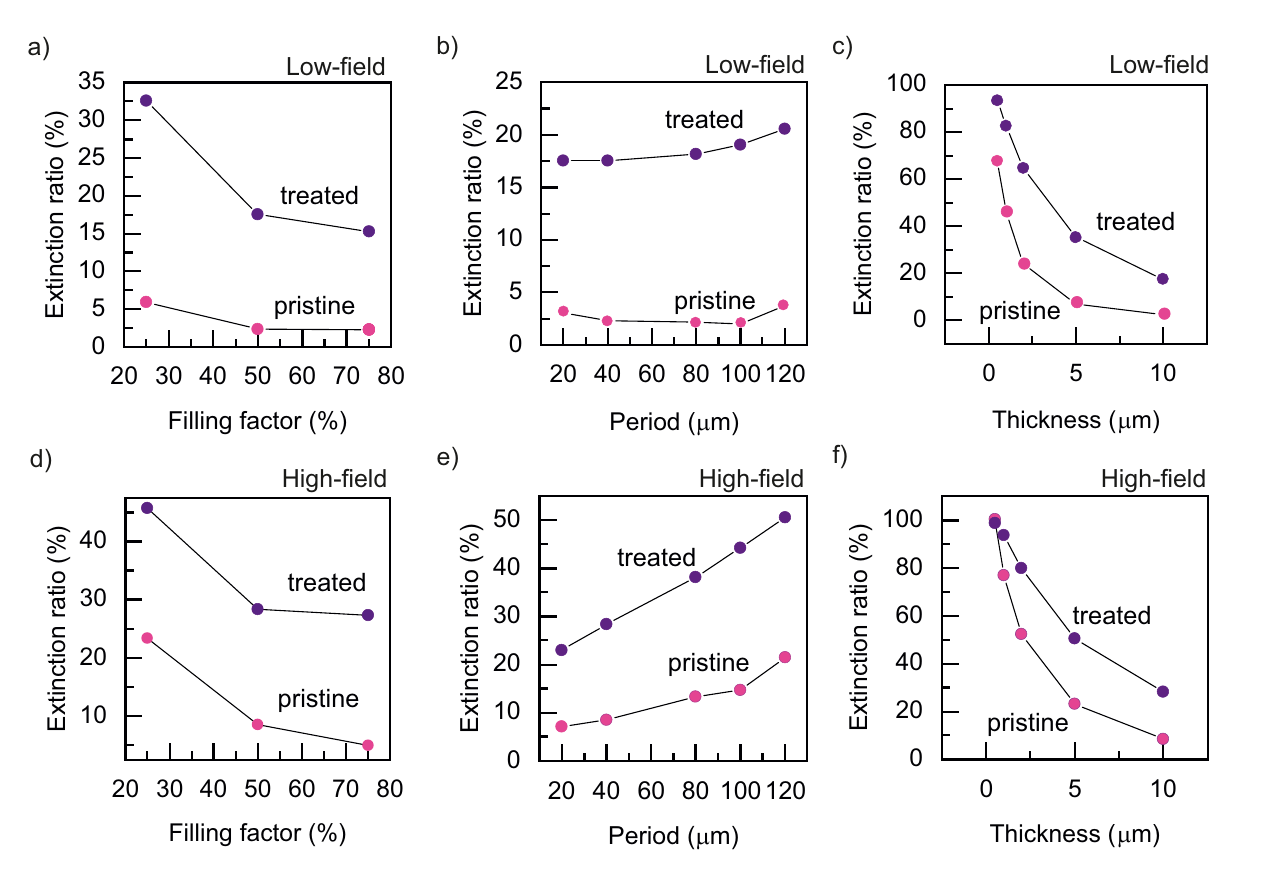}
  \caption{(a) Simulated extinction ratio at 1 THz as a function of filling factor ($w/s$) (a, d), period of the polarizator ($w+s$) (b, e) and its thickness (c, f) at low and high applied electric fields at the top and bottom.}
  \label{FIG:Fig5}
\end{figure*}

\if The spectra of effective dynamical photoconductivity of the samples in the THz range are shown in Fig. \ref{FIG:Fig5}c,d. These spectra are calculated for three different pump--probe delay times to show the relaxation of conductivity to the non--pumped state. As it discovered earlier, the photoconductivity of the SWCNT films tends to be generally negative. There are several explanations of this phenomenon, which are mainly based on the reduction of the conductive charge carriers due to the formation of bound states and the subsequent annihilation of the latter. It was shown in literature that the Lorentzian component remains unchanged by optical pumping, because it is not susceptible to heating. \fi
 
Furthermore, as illustrated in Fig.~\ref{FIG:SIFig3}, there is no substantial alteration in the spectral shape of the real part of the photoconductivity for 10, 60, and 90 s plasma--treated samples. Meanwhile, it is observed that the plasmon frequency shifts towards higher values with an increasing density of defects. This behavior supports the hypothesis that the photoconductivity may exhibit non--plasmonic characteristics. To describe intraband response under illumination, we changed the Drude part of the model. The calculated spectra of photoconductivity were fitted by the difference of two Drude terms, where the first one corresponds to the ON state (optically pumped sample), and the second term -- to the OFF state (no optical pumping). The results are depicted in Fig.~\ref{FIG:Fig5}c,d and ~\ref{FIG:SIFig3}a,b with solid (real part) and dashed lines (imaginary part). The best fitting parameters of the Drude terms with the unchanged Lorentz term for the ON and OFF states were subtracted to define the behavior of the gradual recovery of the parameters (Fig.~\ref{FIG:SIFig3}c,d and Fig.~\ref{FIG:SIFig3}a,b). The increase in scattering rate for plasma--treated SWCNTs in comparison to pristine ones is due to the increased defect density. The consequent recovery of the parameters with pump--probe delay   are evident in Fig.~\ref{FIG:SIFig3}c,d.

\if We qualitatively described the measured conductivity spectra using the semi-classical transport equation, where electronic temperature was employed as a varying parameter. This variation leads to changes in the energy-dependent scattering rate. In SwCNTs and graphene, two primary mechanisms significantly contribute to the overall scattering rate. First, long-range scattering, arising from trapped charge impurities, is inversely proportional to the electron's energy ($\gamma \sim 1/E$). Second, short-range scattering, which affects hot electrons and is caused by neutral impurities like defects, dislocations, and substitutions, is directly proportional to the electron's energy ($\gamma \sim E$) \cite{kar2014, Razavipour2015}. The total scattering time and consequently the scattering rate can be expressed as follows:

\begin{equation}
\frac{1}{\tau}=\frac{1}{\tau_{def}}+\frac{1}{\tau_{Coul}} = A\cdot E+\frac{B}{E},
\end{equation}

\noindent where $A$ and $B$ corresponding weight coefficients.
\fi

\begin{figure*}[t!]
  \centering\includegraphics[scale=0.6]{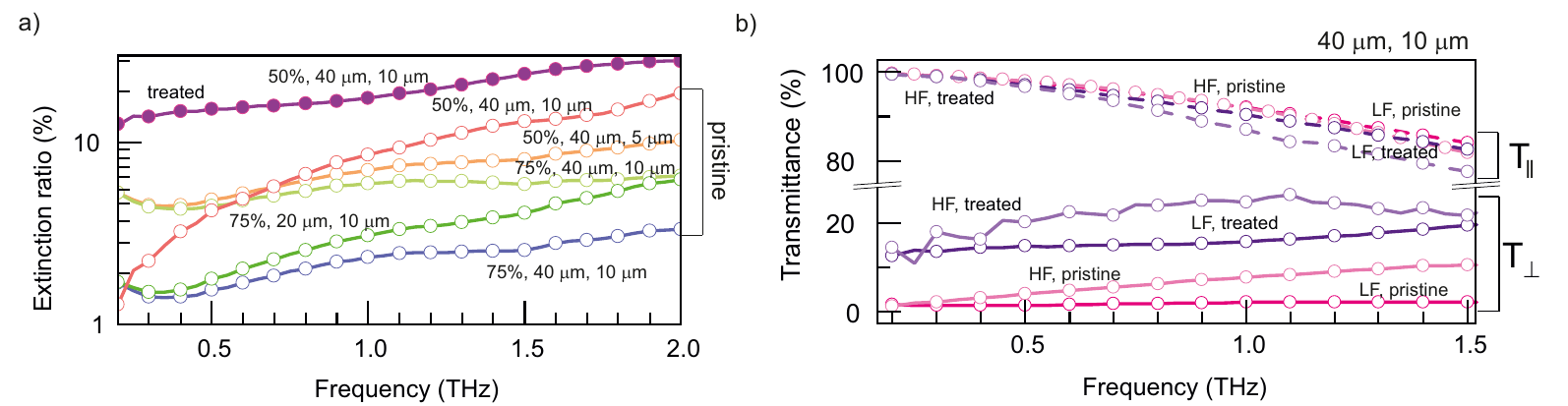}
  \caption{(a) Simulated extinction ratio of SWCNT film polarizer over a broad spectral range at different parameters. Captions denote the filling factor, period and thickness, respectively. (b) The transmission spectra in parallel and perpendicular directions calculated by substituting conductivity presented in Fig.~\ref{FIG:Fig3}. LF and HF stand for low- and high--field transmittance, correspondingly.}
  \label{FIG:Fig6}
\end{figure*}

\if The application of Eqs. 1-4, which assume two primary scattering scenarios, to the experimental data is illustrated in Fig.~\ref{FIG:Fig3}. The derived scattering rates are depicted in Fig. \ref{FIG:Fig4}(b)-(d). For the pristine sample, the best description of the observed data was achieved by using nearly equal weight coefficients for both short and long-range disorder scattering. Other potential terms and consequent scattering mechanisms contributed negligibly and are therefore not discussed here. To assess the validity of the thermodynamic model, we applied it to a high-defects sample. As shown in Fig. \ref{FIG:Fig4}(d), the scattering rate is predominantly governed by the linear term, which corresponds to short--range scattering induced by plasma--introduced defects.

In order to analyze the spectra of conductivity of samples in high-power THz field, we used Ohm’s law $j(t)=\int_0^\infty\sigma(\tau) E(t-\tau)d\tau$, where $j(t)$ is the induced current in the sample. To calculate this current, one may use the Boltzmann approach taking into account processes, which are responsible for scattering of charge carries and affect the momentum relaxation time. These processes may include carrier-carrier collisions, carrier-phonon interactions, carrier-impurity collisions, emergence of excitons, plasmons and etc. Moreover, the studied composites consist of both metallic and semiconducting nanotubes, which requires separate consideration as well as including jumping of hot carriers into higher zones.

\iffalse The previous results from literature show that the scattering in metallic nanotubes might be almost neglected, if the size of the potential barrier, created by impurities, is significant compared to the lattice constant. This limit is a case of the current study: neither in one tube, nor in their massive scattering in metallic NTs is significant. Also in this research the terms of carrier-phonon and carrier-carrier interactions in semiconducting NTs were neglected due to …(написать что-то о времени). \fi

\fi

\subsection{Polarizer performance}
Our investigation of the effect of defects on conductivity under varying electric field strengths provides valuable insights for estimating the performance of THz optoelectronic devices. One such application is the emergent polarizer demonstrated using nanomaterials \cite{Li2020}. We conducted finite-difference time-domain simulations of wire grid polarizers, incorporating the obtained conductivity into the modeling. It is worth noting that it is not obvious at first glance how the sheet conductivity of the randomly oriented nanotubes is affected by the presence of defects. Indeed, the conductivity of metallic nanotubes can only be lessened by impurities, whereas the conductivity of semiconducting nanotubes might be enhanced due to the intermediate defect levels. The parameters varied in the wire-grid polarizer simulations included the strength of the THz electric field (low or high field strengths), defectiveness, filling factor (the ratio between the 'open' and 'closed' part for radiation within one period of a grid), period of the grid, and the thickness of carbon nanotube stripes. 

One of important performance characteristics of polarizers is the extinction ratio (ER), defined as the ratio between the perpendicular and parallel components of the transmitted light (see Fig.~\ref{FIG:Fig6}). Our investigation revealed that by increasing the filling factor, while maintaining a constant period and thickness, the ER gradually decreased. Interestingly, for defective SWCNTs treated with plasma for 90 seconds, the extinction ratio is still low enough, indicating that the device can perform effectively even in the presence of defects. Conversely, we observed a decrease in ER with a reduction in the period and an independent increase in thickness. Nevertheless, the polarizer continued to exhibit superior properties in pristine samples. Notably, we found that the saturation of the ER at thicknesses of 5 and 10 µm for both pristine and defective nanotubes corresponded to the matching of these thicknesses to the skin depth in the low-field regime. As the conductivity decreased at high electric fields, the skin depth effectively doubled, leading to the necessity of utilizing thicker SWCNT wires in designing and implementing the device. This consideration is crucial for optimizing the performance of the polarizer under varying operational conditions.

To further investigate the performance of SWCNT--based polarizers, we calculated the extinction ratio (ER) across a broad spectral range. We observed that the ER increases with frequency due to the lower conductivity of both pristine and defective SWCNT samples. Among all tested parameters, the lowest and most uniform ER was observed at a 75\,\% filling ratio, with a 40 $\mu$m period and a 10 $\mu$m thickness.
The low ER is attributed to the minimal transmittance in the perpendicular direction, which is only a few percent, compared to nearly 100\,\% transmittance in the parallel direction for pristine samples. As anticipated, the ER was higher for defective SWCNT samples, primarily due to a 10\,\% to 20\,\% transmittance in the perpendicular direction. All in all, this section demonstrates how the results of this research might be applied for the engineering a tunable THz device with several working modes. The advantages include the improved extinction ratio by doping with oxygen defects, wide bandwidth with required performance (up to 2 THz). On the other hand, the thin metal films may possess the same values of extinction ratio with less thickness. 

\section{Conclusions}
We present a qualitative, systematic study on the impact of defects induced by the oxygen plasma on the electronic transport properties of SWCNTs, utilizing both low- and high-field time-domain THz spectroscopies. Our investigation monitors the conductivity changes over a wide frequency range and varying electric field strengths. With the increase of the field strength, we observed a decrease in transmission, which we attributed to the thermalization of electrons within the band. Consequently, the conductivity of SWCNTs under a high applied field can be described by the Drude transport equation. Our analysis reveals that the conductivity is predominantly influenced by two contributions: scattering rate exhibits complex $A/E+BE$ type behavior as a function of peak pulse field $E$. In plasma-treated samples, the scattering rate grows monotonically, as opposed to pristine samples. The introduction of defects results in the decrease of the photo-conductivity with the consequent decrease of the lifetimes. Furthermore, we present finite--difference time--domain modeling of a THz wire grid polarizer based on the conductivity spectra of samples subjected to varying field strengths. This modeling, along with our findings, can significantly aid in the design of THz polarizers utilizing SWCNTs with varying levels of defects. This alteration improves the performance of a device, and together with the right choice of optimizable geometry parameters allows to construct wide band light polarizers in the THz range.

\section*{Acknowledgments}
Authors acknowledge the RSF project № 25--79--30006.

%\subsection{Author Contributions}

%\subsection{Conflict of Interest}
%The authors declare no conflict of interest.

\section*{Declaration of competing interest}
The authors declare that they have no known competing
financial interests or personal relationships that could have
appeared to influence the work reported in this paper.

%\section*{References}
\bibliography{mybibfile}

\clearpage

\beginsupplement

\begin{figure*}[t!]
  \centering\includegraphics[scale=0.6]{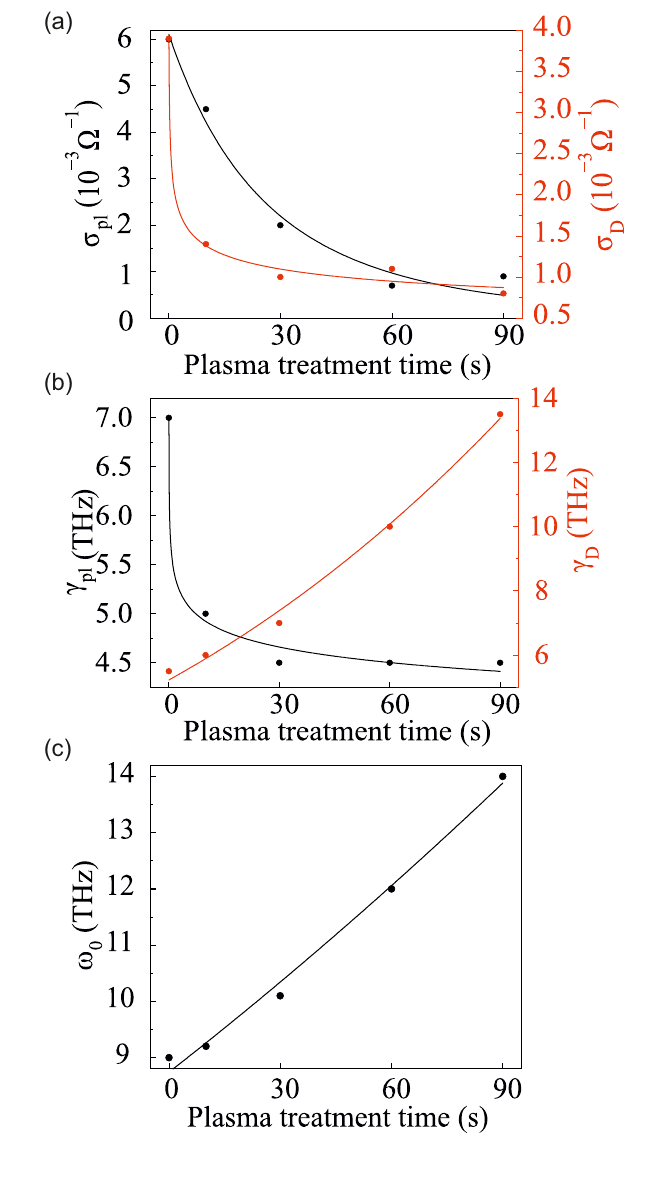}
  \caption{The parameters extracted through fitting the THz optical conductivity spectra  (Fig. \ref{FIG:Fig3}, Main text) with eq (1). a) Plasmon conductivity $\sigma_{pl}$ at the resonance frequency ($\omega$ = $\omega_{0}$, and Drude conductivity $\sigma_D$ in the DC limit. b) phenomenological
scattering rates for the plasmon and free election response, $\gamma_{pl}$ and $\gamma_{D}$, respectively. c) The change of the resonant frequency over plasma treatment time.}
  \label{FIG:SIFig1}
\end{figure*}

\begin{figure*}[t!]
  \centering\includegraphics[scale=0.6]{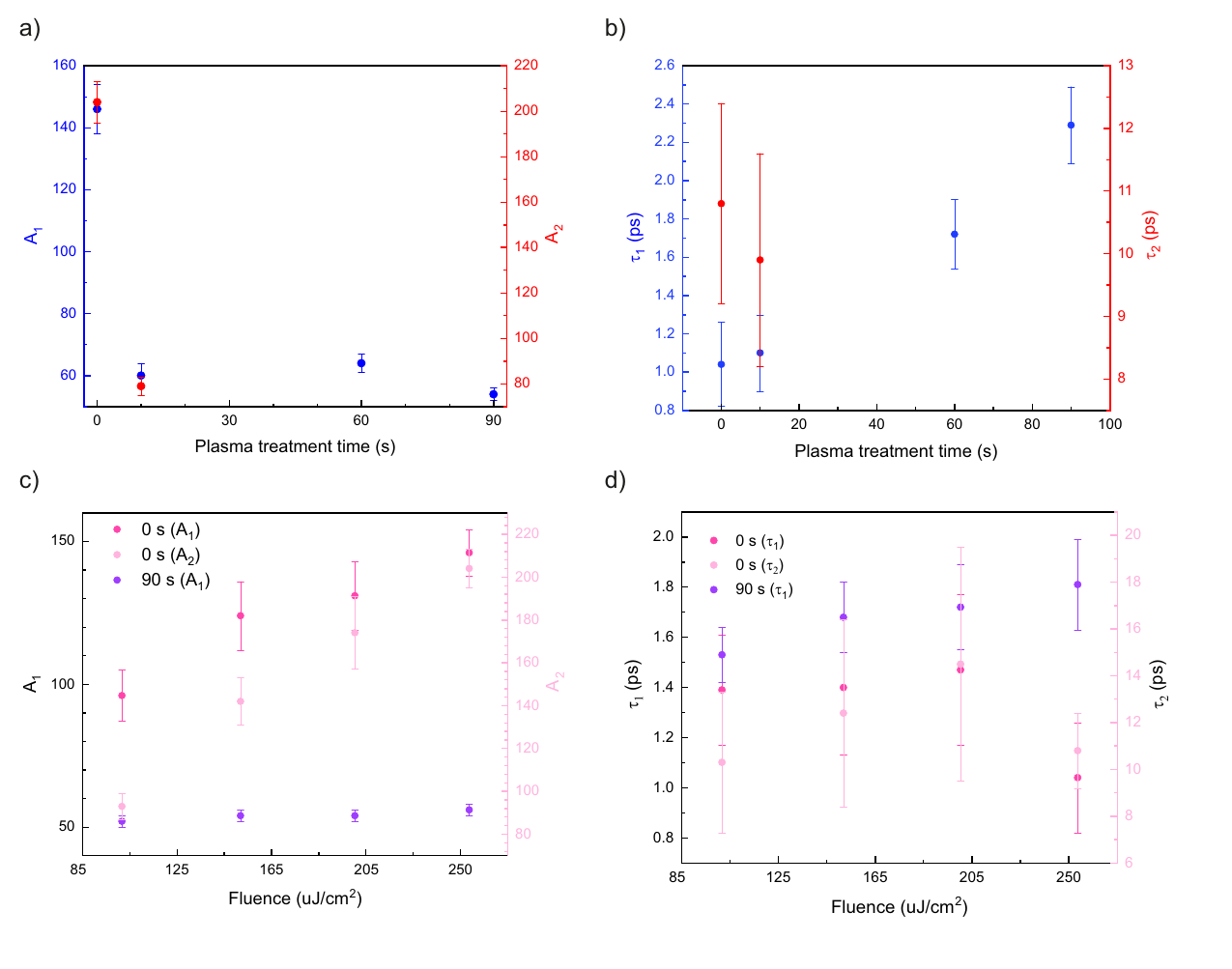}
  \caption{Parameters obtained by fitting by with a bi--exponential decay function: amplitude (a) and relaxation times (b) for fast and slow component as a function of plasma treatment; amplitude (c) and relaxation times (d) for pristine and 90 s treated sample as a function of fluence.}
  \label{FIG:SIFig2}
\end{figure*}

\begin{figure*}[t!]
  \centering\includegraphics[scale=0.6]{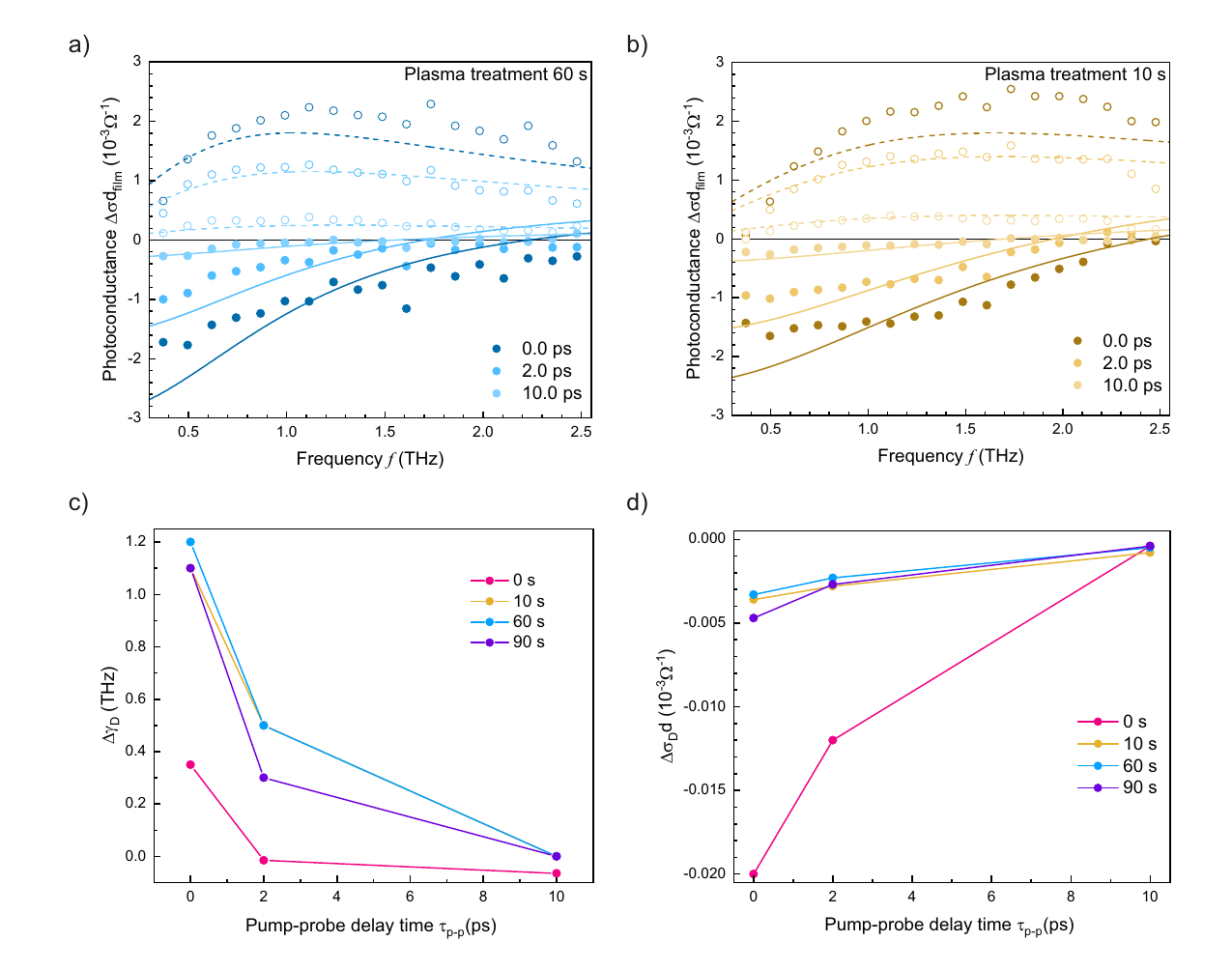}
  \caption{The spectra of photoconductance for plasma-treated during 10 (a) and 60 s (b) samples at different pump-probe delay times. Filled and empty dots are the calculated real and imaginary parts extracted from measurements, solid and dash curves are fits of the real and imaginary parts, correspondingly. The parameters of the photoconductivity fit obtained using the change of the Drude part of the conductivity, scattering rate (c) and conductance (d).}
  \label{FIG:SIFig3}
\end{figure*}

\clearpage

TOC: 

\begin{figure*}[t!]
 \centering\includegraphics[scale=1]{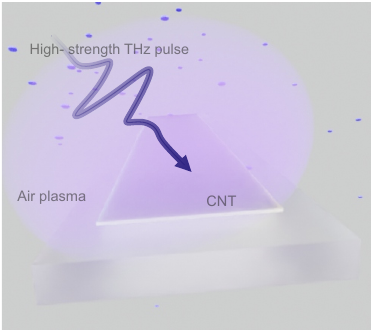}
  \caption{For table of contents.}
  \label{FIG:TOC}
\end{figure*}

\end{document}